\documentclass[11pt]{article}
\usepackage{fullpage}

\begin{document}
\title{\textbf{The sedimentation of flexible filaments: Trajectories, particle clouds and a buckling instability}}
\author{\textsc{Harishankar Manikantan\textsuperscript{1}, Lei Li\textsuperscript{2}, David Saintillan\textsuperscript{1},}\\\textsc{and Saverio E. Spagnolie\textsuperscript{2}} \\ \\ \small
\textsuperscript{1}Department of Mechanical Science and Engineering, University of Illinois at\\ \small Urbana-Champaign, 1206 W. Green Street, Urbana, IL 61801, USA \\ \small
\textsuperscript{2}Department of Mathematics, University of Wisconsin-Madison, 480 Lincoln Drive,\\ \small Madison, WI 53706, USA} 
\date{}

\maketitle

\begin{abstract}
In this fluid dynamics video we explore an array of different possible dynamics for a flexible filament sedimenting in a viscous fluid. The time-dependent shapes and trajectories of the filament are determined analytically and numerically by balancing viscous, elastic and gravitational forces in a slender-body theory for zero-Reynolds number flows. The dynamics are characterized by a single dimensionless elasto-gravitation number. The video shows the process of filament relaxation and reorientation, the formation of particle clouds, and finally a buckling instability.\\
\end{abstract}

\large
\noindent \vspace{2mm}
\textbf{Introduction}
\normalsize

The shapes and dynamics of elastic filaments in viscous fluids are of critical importance in a wide range of biological and technological processes. However, even the study of relatively simple systems can be difficult, due to the complex interactions between elasticity and nonlocal fluid-structure interactions. In a recent paper, the present authors investigated such a fundamental problem, the sedimentation of a single elastic filament in a viscous fluid [1]. A mathematical model was developed by balancing viscous, elastic, and gravitational forces in a slender-body theory for zero-Reynolds number flows. The dynamics of the filament are characterized by a single dimensionless elasto-gravitation number that compares the elastic forces acting on the filament to the gravitational force:
\[\beta=\frac{\pi E a^4}{4 F_G L^2},\]
where $E$ is the elastic modulus of the filament, $a$ is an average cross-sectional radius, $L$ is the filament length, and $F_G$ is the total gravitational force acting on the body. Numerical and analytical approaches are used to study the filament behavior in a number of situations, with excellent agreement between the two [1]. This video illustrates three prominent features of the filament dynamics, which we now describe.\\
\newpage
\large
\noindent \vspace{2mm}
\textbf{1. Particle trajectories and reorientation}
\normalsize

In the first part of the video, numerical simulations are used to show that sedimentation causes weakly flexible filaments to reorient and eventually reach a terminal settling shape. Two non-interacting filaments (with $\beta=0.02$) are initially placed at the origin at angles $\pi/4$ and $\pi/64$ with respect to gravity, and are allowed to sediment freely. The instantaneous shape and orientation are tracked along with the position of the center of the filaments, and the time-dependent trajectories and filament shapes are illustrated. Analytical expressions for these quantities have been determined using a multiple-scale asymptotic expansion in the large $\beta$ regime [1]. We proceed to show the terminal sedimenting shapes over a range of the elasto-gravitation number, $\beta$. These shapes are found (and can be shown analytically) to be self-similar over a broad range of the elasto-gravitation number. When $\beta$ is sufficiently small, large filament deformations lead to a horseshoe-like terminal shape.\\

\large
\noindent \vspace{2mm}
\textbf{2. Particle clouds}
\normalsize

The complicated sedimentation trajectories shown in the first part of the video are a consequence of the coupling between the filament shape and its translational and rotational velocities. The individual filament trajectories suggest an interesting possibility for the spreading dynamics for a collection of flexible filaments, which we model as non-interacting in the dilute limit. The second part of the video shows the sedimentation of a collection of filaments, initially at the same location but with different orientations with respect to gravity, with $\beta=0.1$. In stark contrast to the dynamics of rigid rods, flexible filaments are restricted to a cloud with a width that can be shown to be proportional to the elasto-gravitation number [1].\\

\large
\noindent \vspace{2mm}
\textbf{3. Buckling Instability}
\normalsize

Finally, it can be shown that the tension acting on a filament backbone when it sediments along its length is compressive in the filament's leading half. This compression leads to a buckling instability if the filament is sufficiently flexible. A linear stability analysis reveals this buckling threshold, as well as the most unstable wavenumbers and their growth rates as a function of the elasto-gravitation number. In the final part of the video, we show this buckling instability by numerical simulations for three different values of the elasto-gravitation number ($\beta=0.0005, 0.0001, 0.0000625$). Just as predicted by our analysis, smaller wavelength perturbations are more unstable for smaller $\beta$ (increased filament flexibility).\\ \\


\large
\noindent \vspace{2mm}
\textbf{References}
\normalsize

\noindent
[1] \textsc{Li, L., Manikantan, H., Saintillan, D., \& Spagnolie, S. E.} 2013 The sedimentation of flexible filaments, \textit{J. Fluid Mech}, to appear (arXiv:1306.4692).


\end{document}